\documentclass[12pt]{article}
\begin{document}
\title{On the Threshold of New Physics?}
\author{B.G. Sidharth\\
International Institute for Applicable Mathematics \& Information Sciences\\
Hyderabad (India) \& Udine (Italy)\\
B.M. Birla Science Centre, Adarsh Nagar, Hyderabad - 500 063
(India)}
\date{}
\maketitle
\begin{abstract}
In May 2008 the long awaited NASA Satellite, GLAST was launched,
which will study cosmic gamma rays. In August 2008, it is expected
that the CERN's Large Hadron Collidor (LHC) will finally become
operationized. This is found to unwield several new particles. All
this has the potential of opening the frontier of a New Physics.
\end{abstract}
\section{Introduction}
Particle Physics has stagnated for over three decades. Undoubtedly
the standard model has been a great success, but only at relatively
low energies less than about $200 GeV$ and it is certainly not the
whole story given the fact that there are eighteen arbitrary
parameters and so on. Moreover the discovery of the neutrino mass in
the late nineties in the Super Kamiokande experiment (a prediction
made earlier by the author \cite{cu} and references therein) is a
definite indication that we need to look beyond the standard model
in which the neutrino has vanishing mass. A very important
shortcoming has been the fact that the Higgs Boson has been elusive
for over four decades.\\
At another level, there have been a few contra events observed in
high energy cosmic rays that hint at corrections to Lorentz symmetry
itself (Cf.\cite{lsv} and other references therein).\\
In May this year, NASA's GLAST Telescope finally lifted off and it
is expected that in the next few years it will study gamma rays and
will shed light on some of these puzzles. So also, the physics
community is looking forward to the operationalization of Large
Hadron Collidor ($LHC$) in August. This will be the most powerful
Particle Accelerator for many years to come. This will undoubtedly
provide a wealth of information over the years. Let us now look at
these two important events and possibilities.
\section{Gamma Rays}
Recently the MAGIC Gamma Ray Telescope in Spain detected a big Gamma
Ray flare in the galaxy Mkn 501. It was observed that the luminosity
of the galaxy doubled in two minutes and that there was a four
minute lag between the arrival times of gamma photons with energy
greater than 10 TeV as compared to 100 GeV photons. This was sought
to be explained by Ellis et al as a Quantum Gravity effect in which
there is a dispersion in the velocities of the photons with respect
to frequency \cite{arxiv}. The authors used the timing of the
photons observed by the MAGIC Gamma Ray Telescope during a flare of
the above galaxy and investigated a vacuum refractive index
$$1 - \left(E_0/E'\right)^n , \, \, n=1,2$$
which might represent the effect inducted by Quantum Gravity. They
found that the peeking of the flare maximized for a Quantum Gravity
mass scale $ \sim 0.4 \times 10^{18}GeV$ or $0.6 \times 10^{11}GeV$
for $n=1,2$. They could get lower limits of respectively $0.26
\times 10^{18}GeV$ or $0.30 \times 10^{11}GeV$ at the ninety five
percent confidence level. The sensitivity of the MAGIC telescope at
these levels of confidence was confirmed by Monte Carlo studies.
This does not rule out the contribution of other possible source
effects. They obtained, finally,
$$c' = c(1 - \frac{E_0}{E'})$$
where $c'$ is the modified velocity of the photons and $E'$ is their
estimate for the Planck energy, which turned out to be about one
percent of the actual value. Our work on a non commutative spacetime
predicts a similar effect. This is due to the modified energy
momentum dispersion relation deduced from theory
\cite{ijtp,uof,tduniv}.\\
Based on our earlier considerations, we can deduce from theory that
the usual energy momentum formula is replaced by $(c = 1 = \hbar)$
(Cf.\cite{uof,ijtp2})
\begin{equation}
E^2 = m^2 + p^2 + \alpha l^2 p^4\label{5He1}
\end{equation}
where $\alpha$ is a dimensionless constant of order unity. (For
fermions, $\alpha$ is positive). To see this in greater detail, we
note that, given a minimum fundamental length $l$, the usual Quantum
Mechanical commutation relations get modified and now become, as
shown a long time ago by Snyder,
\begin{equation}
[x,p] = \hbar' = \hbar [1 + \left(\frac{l}{\hbar}\right)^2 p^2]\,
etc,\label{5He2}
\end{equation}
$$[x,y] = O(l^2) etc.$$
where we have temporarily re-introduced $\hbar$ (Cf. also
ref.\cite{bgsust}). (\ref{5He2}) shows that effectively $\hbar$ is
replaced by $\hbar'$. Interestingly (\ref{5He2}) is Lorentz
invariant for any minimum length $l$. In our usual (commutative)
spacetime, the left side of (\ref{5He2}) would vanish for
coordinates $x$ and $y$. Strictly speaking relations (\ref{5He2})
would hold in Quantum Gravity approaches and even M theory. So, from
(\ref{5He2}), we get, with the new $\hbar'$,
$$E = [m^2 + p^2 (1 + l^2 p^2)^{-2}]^{\frac{1}{2}}$$
So the energy-momentum relation leading to the Klein-Gordon
Hamiltonian is now given by, from the above,
\begin{equation}
E^2 = m^2 + p^2 - 2l^2 p^4,\label{5He3}
\end{equation}
neglecting higher order powers of $l$.\\
For Fermions the analysis can be more detailed, in terms of Wilson
lattices \cite{mont}. The free Hamiltonian now describes a
collection of harmonic fermionic oscillators in momentum space.
Assuming periodic boundary conditions in all three directions of a
cube of dimension $L^3$, the allowed momentum components are
\begin{equation}
{\bf q} \equiv \left\{q_k = \frac{2\pi}{L}v_k; k = 1,2,3 \right\},
\quad 0 \leq v_k \leq L - 1\label{4.59}
\end{equation}
(\ref{4.59}) finally leads to
\begin{equation}
E_{\bf q} = \pm \left(m^2 + \sum^{3}_{k=1} a^{-2} sin^2
q_k\right)^{1/2}\label{4.62}
\end{equation}
where $a = l$ is the length of the lattice, this being the desired
result. (\ref{4.62}) shows that $\alpha$ in (\ref{5He1}) is
positive. We have used the above analysis to indicate that in
the Fermionic case, the sign of $\alpha$ is positive.\\
A rigid lattice structure imposes restrictions on the spacetime -
for example homogeneity and isotropy. Such restrictions are not
demanded by fuzzy spacetime, and we use the lattice model more as a
computational device (Cf. ref.\cite{uof} and \cite{lsv}). This leads
to a modification of the Dirac and Klein-Gordon equations at ultra
high energies (Cf.ref.\cite{uof,ijtp2,ijmpe2}). It may be remarked
that proposals like equation (\ref{5He1}) have been considered by
several authors though from a phenomenological point of view (Cf.
refs.\cite{kif}-\cite{cam}). Our approach however, has been
fundamental rather than phenomenological in that we start with
(\ref{5He2})
and deduce (\ref{5He3}).\\
Using (\ref{5He3}), we can easily deduce that
$$E^2 = E^2_0 (1 - \frac{E^2_0}{E^{''2}})$$
This gives
\begin{equation}
c^{'2} = c^2(1 - \frac{E^2_0}{E^{''2}}\label{e6}
\end{equation}
where $E^{''}$ is the actual Planck energy. On the other hand from
the equation above of Ellis et al. we get
\begin{equation}
c^{'2} = c^2(1 - \frac{E^2_0}{E^{''2}} \cdot 10^4)\label{e7}
\end{equation}
A comparison of (\ref{e6}) and (\ref{e7}) shows that though the
approach of Ellis et al., is interesting, their limits do not
exactly reproduce our exact relation (\ref{e6}).\\
Finally, it may be reiterated that the just launched GLAST Gamma Ray
Telescope of NASA would almost certainly shed further light on the
matter.
\section{New Particles}
The physics community is expectantly awaiting the Large Hadron
Collidor ($LHC$), as we are certain to get a wealth of new data with
the potential to even transform Particle Physics. It is known that
the standard model is in excellent agreement with experiments at
energies around or less than $200 GeV$. This is a theory of weak and
electromagnetic interactions. However as noted in Section 1, this is
not the whole story (\cite{uof,tduniv} and references therein).
There are some important unanswered questions. These include the
question why the elementary particles have the masses which they are
experimentally known to have, that is the question of the mass
spectrum. There is also the case of the elusive Higgs Boson which is
required in the theory for generating the mass of the particles. The
Higgs Boson, despite several attempts, has not turned up. Could the
$LHC$ discover the Higgs Boson? Equally interesting, what if the
Higgs
Boson were not discovered even by the $LHC$?\\
Then there are subtler issues. For instance the dominance of matter
over anti matter in the universe--this observational fact requires
CP Violation, which indeed was observed way back in 1964 in the
decay of $K$ meson. Though this is in agreement with the standard
model, it does not still explain the observed matter-anti matter
ratio. In other words we need further sources of CP Violation.
Incidentally this could also result from the Lorentz Symmetry
Violation alluded
to in Section 2.\\
Another important issue is that of Supersymmetry (SUSY). Though this
bas been an elegant theory which has even threatened to solve the as
yet unsolved problem of the unified description of General
Relativity and Quantum Theory, the fact is that it predicts a whole
range of Supersymmetric particles which have not yet been detected.
If some or all of these particles are detected by the $LHC$, that
would be a major headway in Particle Physics and Quantum Gravity.\\
All this including the question of a Neutrino Mass would constitute
what has been come to be known as physics beyond the standard model.
This would also include the new paradigm of Dark Energy, which was
indirectly detected through the acceleration of the universe by an
observation of distant supernovae in 1998. Indeed the author's 1997
model had predicted this new cosmological scenario (Cf.ref.\cite{cu}
and references therein). So it is no
wonder that so many hopes are pinned on the $LHC$.\\
Finally there is also the related problem of Scale Invariance. This
is a well studied problem, at least mathematically, in which we
require that the physical laws remain unchanged if the length or
energy scale of the problem is multiplied by some factor. However it
has not been found as yet in the standard model. It appears that we
may require what are called un particles for this to be the case
\cite{akgiri}.\\
In any case the $LHC$ is bound to come up with as yet unknown
particles. It is interesting that the author's 2003 mass spectrum
formula viz.,
\begin{equation}
m_P = m \left(n + \frac{1}{2}\right) m_\pi\label{4Ge4}
\end{equation}
gives the mass of all known elementary particles. In the above
equation $m_P$ is the mass of the elementary particle in question
and $m_\pi$ is the mass of the $\pi$ meson and $m$ and $n$ are
positive integers. It is derived based on the $QCD$ potential
\cite{bgshadron,uof,tduniv}. It gives the mass of all the known
elementary particles with an error of about three percent or less.
The subsequently discovered $Ds (2317)$ and the as yet unconfirmed
$1.5 GeV$ Pentaquark as also the $Ds (2632)$ and the more recent
$4.43 GeV$ meson, the so called $Z$ charged mesons are also
described by the above formula. It would be interesting to see if
the new particles which are bound to be discovered by $LHC$ would
obey the formula (\ref{4Ge4}).

\end{document}